\documentclass[preprint,12pt]{elsarticle}

\usepackage{epsfig}
\usepackage{subfigure}
\usepackage{float}

\usepackage{amsmath}
\usepackage{amsfonts}
\usepackage{amssymb}
\usepackage{amsbsy}
\usepackage{bm}

\newcommand{\lla}{\left\langle}
\newcommand{\rra}{\right\rangle}

\journal{Applied Numerical Mathematics}

\begin{document}

\begin{frontmatter}



\title{Tethered flexible polymer under oscillatory linear flow}


\author[a]{A. Lamura}
\ead{antonio.lamura@cnr.it}
\affiliation[a]{organization={Istituto per le Applicazioni del Calcolo, CNR},
            addressline={Via Amendola 122/D}, 
            city={Bari},
            postcode={I-70126}, 
            country={Italy}}
\begin{abstract}
The non-equilibrium structural and dynamical properties of a flexible polymer 
tethered to a reflecting wall and subject to oscillatory linear flow are 
studied by numerical simulations. Polymer is confined in two dimensions and
is modeled as a bead-spring chain immersed in a fluid
described by the Brownian multiparticle collision dynamics.
At high strain, the polymer is stretched along the flow direction following
the applied flow, then recoils at flow inversion before flipping and elongate
again. When strain is reduced, it may happen that the chain recoils without
flipping when the applied shear changes sign. 
Conformations are analyzed and compared to stiff polymers
revealing more compact patterns at low strains and less stretched
configurations at high strain. 
The dynamics is investigated by looking at the center-of-mass
motion which shows a frequency doubling along the direction
normal to the external flow. The center-of-mass correlation function is 
characterized by smaller amplitudes when reducing bending rigidity.
\end{abstract}


\begin{highlights}
\item Presence of three conformational regimes when varying external strain
\item Flexible chains are more compact at low strains and less
  stretched at high strains with respect to stiff polymers
\item Smaller amplitudes in the center-of-mass autocorrelation function when
      reducing bending rigidity
\end{highlights}

\begin{keyword}
Mathematical modeling \sep numerical simulation \sep flexible polymer
\sep oscillatory flow



\end{keyword}

\end{frontmatter}


\section{Introduction}
\label{intro}

The behavior of individual semiflexible polymers under the action of mechanical 
solicitations has been largely explored in the past. The interest
mainly prompts from the response to external stimuli 
of biological filaments, such as DNA and polypeptides, whose rigidity impacts
on their physiological functions.
Different techniques are available to manipulate single polymers
which can be stretched either by forces or fields. In the former case,
tension is applied on a filament fixed at one end by using optical (or magnetic)
tweezers
\cite{smit:96-2,ashk:97,stri:99,mollo:02,goss:02}
or atomic force microscopy 
\cite{czaj:98,alle:03}.
Another possibility to study the 
stretching dynamics is provided by the use of an external
field. In this case, it is possible to use a flow field
\cite{trah:09,gade:10,thur:11,hsie:11,hu:12,wang:12,faru:16}
or an electric (or
magnetic) field
\cite{schw:93,stri:96,stri:00}.
Complementary to experimental techniques are theoretical models 
\cite{mark:95,hofm:00,dai:03,phil:08,gros:11}
as well as numerical simulations
\cite{liu:89,pier:95,knud:96,hur:00,jend:02,hsie:04,liu:04,grat:05,pami:05,schr:05,shaq:05,cela:05,ryde:06,ripo:06,jose:08,send:08,zhan:09,he:09,wink:10,huan:10,huan:11,lamu:12,huan:12}
which
contributed in pointing out novelties in 
dynamics, conformations, and rheological behavior.

The large majority of studies about free and tethered filaments
under flow consider steady shear and extensional flows.
These conditions are sufficient in order
to gain insight into linear rheological behavior.
Oscillatory shear flow  can also be used in order to
scrutinize the properties
of complex fluids. In particular,
large-amplitude oscillatory shear (LAOS) \cite{hyun:11}
offers the chance
of investigating nonlinear aspects obtaining a stress response
which is typically of complex shape, and no longer sinusoidal,
since the conformational properties of single polymers are greatly affected
\cite{zhou:16}. The structure and dynamics of pinned
semiflexible polymers subject
to LAOS have been numerically investigated \cite{lamu:19,lamu:21}.

The behavior of flexible filaments has been so far much less investigated.
Experiments allowed the determination of the force-extension curve of
flexible polymers such as poly(methylacrylic acid) finding good agreement
with the description provided by the freely-jointed chain
model \cite{kout:97}.
Tethered flexible polymers under steady shear 
\cite{delg:06,iban:09,anand:17}
and free ones under LAOS \cite{bona:23} have been studied in three dimensions
by using numerical models. 
Two-dimensional systems received less attention \cite{lado:00,manc:12}
despite the fact that novel nonequilibrium dynamical properties
may be expected when a polymer is 
strongly adsorbed on a substrate \cite{maie:02}.

In the present paper, 
a flexible Rouse filament, tethered at one end on a reflecting wall, is 
immersed in a Brownian solvent and exposed to an oscillatory shear flow
in two dimensions. The conformations and the dynamical behavior are numerically
investigated.
In the case of chains with negligible bending rigidity, 
hydrodynamic interactions are able to 
influence dynamics. Such interactions 
are expected to be of minor importance
for rather 
elongated configurations \cite{lado:00} but can accelerate the motion especially
in the coiled state during flow inversion. Despite of this,
hydrodynamics is not considered
in this study in order to point out the relevant 
features that are induced by the oscillatory shear flow itself since
no study is available in the literature up to now.
The chain is modeled as a worm-like filament with vanishing small
 bending rigidity.
Excluded-volume interactions, which are essential in two dimensions,
are explicitly taken into account. The polymer
is immersed in a heat bath which is
numerically implemented by using the Brownian multiparticle-collision (B-MPC)
method \cite{kiku:03,ripo:07,kapr:08,gomp:09}. 
Conformations reveal the existence of three regimes when varying the strain,
which is expressed as the ratio of the shear rate to the shear frequency.
At low strains, polymers behave as at equilibrium and no remarkable shear
effect can be detected. In the opposite limit of high strain, polymers follow
the imposed flow alternatively elongating on both sides after flipping during
flow inversion. More interesting is what observed at intermediate values
of strain when flipping may be halted. This feature
favors more coiled configurations
characterized by larger bond angles with respect to 
semiflexible polymers. The internal dynamics is investigated by 
looking at the autocorrelation function
of the center of mass (COM) whose motion is periodic,
displaying the same frequency of the imposed shear along the flow
direction and a double frequency along the normal direction. 
Negligible bending rigidity 
promotes the reduction of the amplitude of the periodic motion 
of the COM along the normal direction and allows a faster
 relaxation, observed by looking at the mode-autocorrelation function.

The mathematical model and its numerical implementation are 
illustrated in Section \ref{model}. The results are 
presented in Section \ref{results} and, finally, the main outcomes
of this study are discussed in Section \ref{concl}.

\section{The model}
\label{model}

A discrete chain of total length $L$, made 
of $N$ beads of mass $M$, is considered in two spatial dimensions.
The first bead, with position index $i=1$, is maintained pinned all the time
at the point $(0,0)^T$ of the Cartesian
space. The filament is confined in the half-plane $y>0$ and
initially aligned along the $y$-direction (shear direction)
with no fixed orientation of the first bond during simulations.
A total potential $U=U_{bond}+U_{bend}+U_{ex}+U_w$, which comprises different
contributions, is introduced. 
In order to preserve on the average the total contour length,
harmonic springs connect consecutive beads
so that the bond potential is 
\begin{equation}\label{bond}
U_{bond}=\frac{\kappa_h}{2} \sum_{i=1}^{N-1}
(|{\bm r}_{i+1}-{\bm r}_{i}|-r_0)^2 ,
\end{equation}
where $\kappa_h$ is the spring constant, ${\bm r}_i=(x_i,y_i)$
is the the position vector of the $i-$th bead
($i=1,\ldots,N$), and $r_0$ is the bond
rest length.
The bending potential
\begin{equation}
U_{bend}=\kappa \sum_{i=1}^{N-2} (1-\cos \varphi_{i})
\label{bend}
\end{equation}
allows us to modify the chain stiffness by varying the bending rigidity
$\kappa$, $\varphi_{i}$ being the angle between
two consecutive bond vectors.
The persistence length $L_p$, which measures the length at which two bonds
become decorrelated,
is related to the rigidity by the relationship
$L_p=2 \kappa r_0/ k_B T$ \cite{wink:94}
where $T$ is the temperature and $k_B$ is
Boltzmann's constant.
In order to avoid overlap of beads, 
the shifted and truncated Lennard-Jones potential
\begin{equation}
U_{ex} =
4 \epsilon \sum_{i=1}^{N-2} \sum_{j=i+2}^{N}
\Big [ \Big(\frac{\sigma}{r_{i,j}}\Big)^{12}
  -\Big(\frac{\sigma}{r_{i,j}}\Big)^{6} +\frac{1}{4}\Big]
\Theta(2^{1/6}\sigma -r_{i,j}) 
\label{rep_pot}
\end{equation}
is considered among non-connected beads where $\epsilon$ is the
interaction energy, 
$r_{i,j}$ is the relative distance between two beads, and
$\Theta(r)$ is the Heaviside function ($\Theta(r)=0$ for $r<0$ and
$\Theta(r)=1$ for $r\ge 0$). 
Finally, the confining wall at $y=0$ is enforced by using the potential 
\begin{equation}
U_{w} =
4 \epsilon  \sum_{i=2}^{N} \Big [ \Big(\frac{\sigma_w}{y_i}\Big)^{12}
  -\Big(\frac{\sigma_w}{y_i}\Big)^{6} +\frac{1}{4}\Big]
\Theta(2^{1/6}\sigma_w -y_i) 
\label{rep_pot2}
\end{equation}
where $y_i$ ($i=2,\ldots,N$) are the bead distances from the wall. 
Newton's equations of motion of beads are integrated by
the velocity-Verlet algorithm with time step $\Delta t_p$
\cite{swop:82,alle:87} so that the positions ${\bm r}$
and velocities ${\bm v}$ of beads
are updated at time $t$ according to the prescription
\begin{eqnarray}
  {\bm r}_{i}(t+\Delta t_p)&=&{\bm r}_{i}(t)+{\bm v}_{i}(t) \Delta t_p
  +\frac{1}{2}{\bm a}_{i}(t) \Delta t_p^2, \;\;\;\;\;\;\;\;\;\; i=2,\ldots,N\\
  {\bm v}_{i}(t+\Delta t_p)&=&{\bm v}_{i}(t)
  +\frac{1}{2}[{\bm a}_{i}(t) + {\bm a}_{i}(t+\Delta t_p)] \Delta t_p , \;\;\;
  i=2,\ldots,N,
\end{eqnarray}
where ${\bm a}_{i}(t)=-\frac{1}{M} \nabla_{{\bm r}_i} U$ ($i=2,\ldots,N$)
are the bead accelerations.

Hydrodynamic interactions are ignored and the polymer is
immersed in a heat bath
which is modeled
by implementing the Brownian version 
\cite{ripo:07,gomp:09} of the multiparticle collision (MPC)
method \cite{ihle:01,lamu:01}.
In this algorithm beads collide with 
a virtual particle of mass $M$
in order to mimic the interaction with a
fluid volume, made of $\gamma$ particles of mass $m$, surrounding the bead.
The linear momentum ${\bm P}(t)=(P_x(t),P_y(t))^T$
of the fluid particle at time $t$ is
a stochastic variable whose components $P_x$ and $P_y$ are randomly sampled
from a Maxwell-Boltzmann distribution.
The variance is $M k_B T$ and is related to the temperature of the heat bath
while the mean takes into account the flow properties.
In the present model, the mean value of the distribution for $P_x$
is $M \dot{\gamma} y_i \sin(\omega t)$ in order to enforce
an oscillating linear flow, on the $i$-th bead located
at a distance $y_i$ 
from the wall,
oriented along the $x$-direction (flow direction) with 
shear rate $\dot \gamma$ and angular frequency $\omega$.
The flow direction is fixed in time and the shear rate is time-modulated
by the oscillatory term.
The mean value of the distribution for $P_y$ is $0$ due to the lack of any
imposed flow along the $y$-direction (shear direction).
Collisions between beads and virtual solvent particles
are executed via the stochastic rotation dynamics
of the MPC method \cite{kapr:08,gomp:09} in the following way.
The relative velocity of a polymer bead at time $t$, with respect to the
COM velocity of the bead and the coupled solvent
particle, is randomly rotated in the $xy$-plane by angles $\pm \alpha$
so that the updated bead velocity is given by
\begin{equation}
  {\bm v}_i(t+\Delta t)={\bm v}_{cm,i}(t)+
  \Omega(\alpha) [{\bm v}_i(t)-{\bm v}_{cm,i}(t)], \;\;\;\;\; i=2,\ldots,N.
\end{equation}
The matrix
\begin{equation}
\Omega(\alpha) = 
\begin{pmatrix}
\cos \alpha  & -\sin \alpha  \\
\sin \alpha & \cos \alpha
\end{pmatrix}
\end{equation}
denotes a stochastic rotation matrix which rotates by an angle of either
$+\alpha$
or $-\alpha$ with probability $1/2$
and the COM velocity is
\begin{equation}
  {\bm v}_{cm,i}(t)= \frac{M {\bm v}_i(t) + {\bm P}(t)}{2 M}, \;\;\;\;\; i=2,\ldots,N .
\end{equation}  
Collisions occur at time intervals $\Delta t$ much longer
compared to $\Delta t_p$.
It can be shown \cite{kiku:03}
that the evolution equation of the MPC model for the
bead takes the form of a discretized Langevin equation 
\begin{equation}
M \frac{{\bm v}_i(t+\Delta t)-{\bm v}_i(t)}{\Delta t}=
- \xi {\bm v}_i(t) + {\boldsymbol \eta}(t), \;\;\;\;\; i=2,\ldots,N
\end{equation}
where
\begin{equation}
\xi=\sum_{n=0}^{+\infty} \frac{e^{-\gamma}\gamma^n}{n!} \frac{m n}{M+m n} \frac{M}{\Delta t} (1-\cos \alpha)
\end{equation}
is the friction coefficient \cite{kiku:03},
and ${\boldsymbol \eta}(t)$ is a stochastic noise with
$< {\boldsymbol \eta}>=0$ and $<{\boldsymbol \eta}(t) \cdot 
{\boldsymbol \eta}(t)>= 4 \xi k_B T/\Delta t$  \cite{kiku:03}.

The selected parameters for simulations are $\alpha=130^{o}$,
$\Delta t=0.1 t_u$, where the time unit is $t_u=\sqrt{m r_0^2/(k_BT)}$,
$M=\gamma m$ with $\gamma=5$,
$\kappa_h r_0^2/(k_B T)=4 \times 10^3$, 
$\epsilon / (k_B T)=1$, $\sigma=r_0$, $\sigma_w=r_0/2$, $N=101$,
and $\Delta t_p=10^{-2} \Delta t$. The large value of
$\kappa_h$ is set to preserve on the average the total
length $L=100r_0$ of the polymer.
The bending rigidity is $\kappa = k_B T$ in order to simulate a flexible
filament with negligible persistence length compared to the polymer length
since it results to be $L_p/L=0.02$.
The end-to-end vector (${\bm R}_e= {\bm r}_N - {\bm r}_1$) relaxation time,
estimated by 
simulations of free polymers, is
$\tau_r \simeq 0.7 \times 10^6 t_u$ \cite{lamu:12}.
$\tau_r$ is defined as 
the time at which the autocorrelation function of the
 end-to-end vector ${\bm R}_e$ decays to $1/e$.
For the set of the
parameters adopted in the manuscript, it results to be $M/\xi
\simeq 1.9 \times 10^{-7} \tau_r$ so that 
the dynamics is close to be overdamped and inertial effects
are negligible for the shown results.
The present setup is characterized by the presence of three time scales:
The polymer relaxation time $\tau_r$, the inverse shear rate $1/\dot\gamma$, and
the shear period $1/\omega$. Their relative ratios allow us to characterize
simulated systems in terms of two dimensionless numbers,
namely, the Weissenberg number
$Wi=\tau_r \dot{\gamma}$ and the Deborah number $De=\tau_r \omega$.
Values in the ranges $0 \leq Wi \leq 2 \times 10^3$ and
$0.1 \leq Wi/De \leq 100$ are accessed where 
the further ratio $Wi/De=\dot{\gamma}/\omega$
is proportional to the strain in a half-cycle.

\section{Numerical results}
\label{results}

The conformations and dynamics of a tethered flexible polymer with
negligible persistence length
under oscillating shear flow are studied and compared to the properties
of semiflexible filaments under the same flow conditions. 
The polymer is 
let equilibrate up to time $t_w = 5 \times 10^6 t_u > \tau_r$.
Data are then registered up to the time $10^{8} t_u \gg \tau_r$ and
averaged over half-periods with positive shear flow. 

\begin{figure}[H]
\begin{center}
\includegraphics*[width=0.45\columnwidth,angle=0]{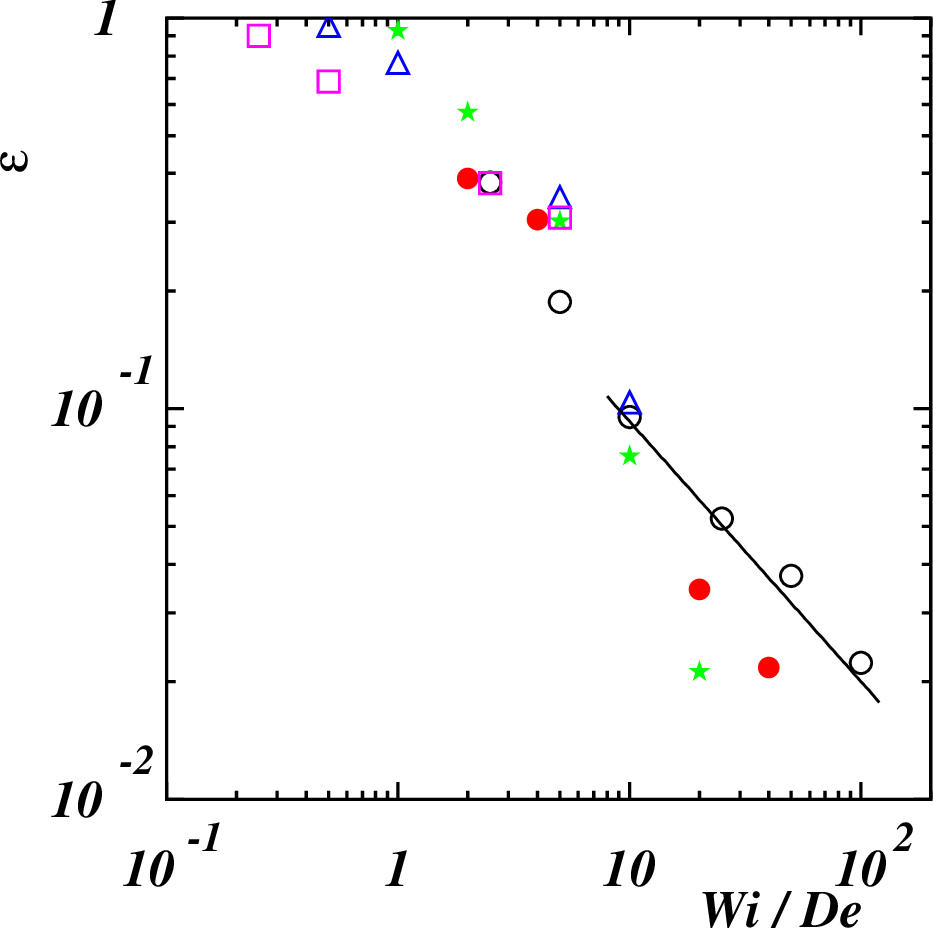}
\caption{Mean deficit length-ratio $\varepsilon$ 
as a function of the strain $Wi/De$
for $L_p/L=0.02$ and
$De=10$ ($\circ$), $25$ ($\bullet$), $50$ ($\star$),
$100$ ($\triangle$), and $200$ ($\square$).
The slope of the full line is $-2/3$.
\label{fig:deficit}
}
\end{center}
\end{figure}

\begin{figure}[H]
\begin{center}
\includegraphics*[width=0.45\columnwidth,angle=0]{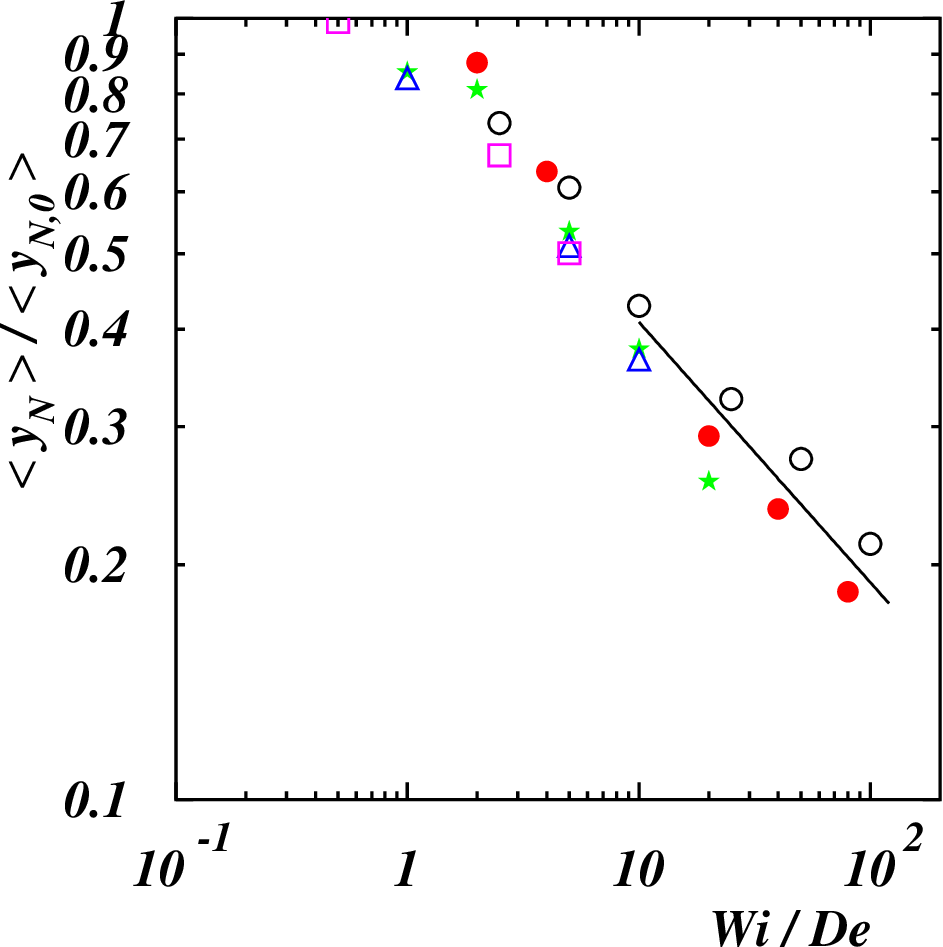}
\caption{Mean polymer height above the wall as a function of the strain $Wi/De$
for $L_p/L=0.02$ and
$De=10$ ($\circ$), $25$ ($\bullet$), $50$ ($\star$),
$100$ ($\triangle$), and $200$ ($\square$).
The solid line indicates
the slope $-1/3$.
$\lla y_{N,0} \rra$ is the mean polymer height at equilibrium.
\label{fig:height}
}
\end{center}
\end{figure}

\subsection{Conformations}

The end-to-end vector ${\bm R}_e={\bm r}_N-{\bm r}_1$ 
is considered to highlight the polymer conformations.
The maximum extension $R_{ex,max}$ of the polymer
along the $x$-axis in a cycle is considered 
to compute
the average deficit-length ratio $\varepsilon=1-\lla R_{ex,max} \rra/L$ which
is shown in Fig.~\ref{fig:deficit} as a function of the strain $Wi/De$.
Data for different values of $De$ collapse on the same curve
when plotted versus strain making possible to identify 
three regimes, referred to as
low strain ($Wi/De \lesssim 1$), intermediate
strain ($1 \lesssim Wi/De \lesssim 10$), and high strain ($Wi/De \gtrsim 10$).
In the first regime, 
the applied flow is not able to stretch remarkably the polymer.
By increasing strain,
the deficit-length ratio decreases sensibly showing
a power-law behavior for high values of strain with an
exponent $-2/3$ for $De=10$,
as predicted for tethered flexible chains
under steady shear \cite{lado:00}. This is at odd with the exponent
$-1/3$ observed 
for semiflexible chains \cite{lamu:21}. 
Higher values of the shear rate would be required in order to confirm
the power-law behavior even for values $De > 10$.
Moreover,
flexible polymers appear to be less stretched compared to stiffer ones
which show smaller values of $\varepsilon$ \cite{lamu:21}.

The presence of the repulsive wall forces polymer stretching normal to the
wall. It is possible to define the polymer height 
as the coordinate $y_N$ of the last bead.
The average values $\lla y_N \rra$ 
are reported in Fig.~\ref{fig:height}.
A scaling regime can be observed
at high strain with dependence $\lla y_N \rra \sim (Wi/De)^{-1/3}$
as for semiflexible polymers \cite{lamu:21},
confirming the prediction valid both for flexible and semiflexible
polymers under steady shear flow when hydrodynamics is neglected
\cite{lado:00}.

\begin{figure}[H]
\begin{center}
\includegraphics*[width=\textwidth,angle=0]{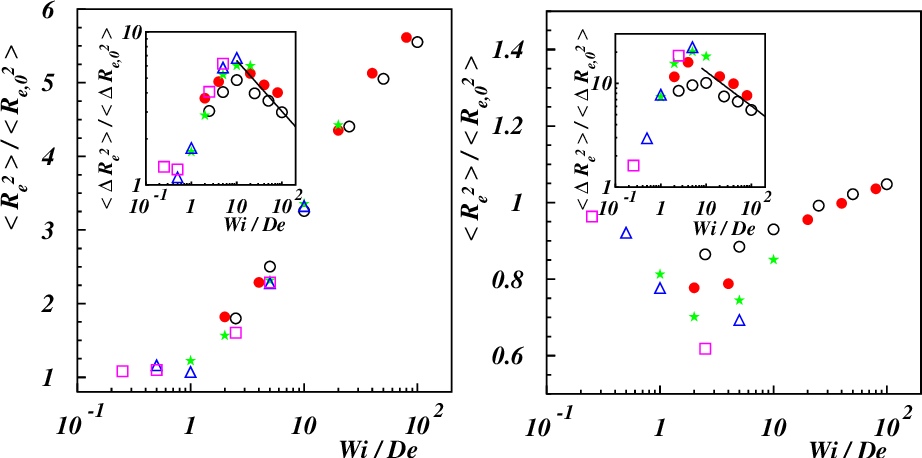}
\caption{Mean square polymer end-to-end distances 
as function of the strain $Wi/De$
for $L_p/L=0.02$ (left), $2$ (right) and
$De=10 (\circ), 25 (\bullet), 50 (\star), 100 (\triangle), 
200 (\square)$. $\lla R^2_{e,0} \rra$ is the mean square 
value of the polymer end-to-end distance at equilibrium.
Insets: Variance of the polymer end-to-end distance 
as a function of the strain $Wi/De$. 
$\lla \Delta R^2_{e,0} \rra$ is the variance of the
polymer end-to-end distance at equilibrium. The straight lines have
slope $-1/3$.
\label{fig:re}
}
\end{center}
\end{figure}
Information about the conformation of the polymer can be gained by looking
in more detail at the behavior of the end-to-end distance whose
mean square values are reported
in Fig.~\ref{fig:re} for $L_p/L=0.02, 2$.
At small strains, $\lla R_e^2 \rra$ is comparable 
to the equilibrium value. When $Wi/De > 1$, the flexible filament is
continuously elongated when increasing the strain,
differently from what happens for the stiffer
chain. In this latter case,
the polymer contracts reaching a minimum in the end-to-end distance
at $Wi/De \simeq 2.5$ and
larger values of the strain favor a slow polymer swelling
in the high strain regime.
The behavior of the variance 
$\lla \Delta R_e^2\rra=\lla R_e^2 \rra -\lla R_e \rra^2$
is qualitatively the same for both the bending rigidities:
It increases first with
strain, reaches a maximum at $Wi/De \simeq 8$, whose height
 depends on the persistence
length, and then decreases with
a power-law with an exponent $-1/3$.

\begin{figure}[H]
\begin{center}
\includegraphics*[width=.45\textwidth,angle=0]{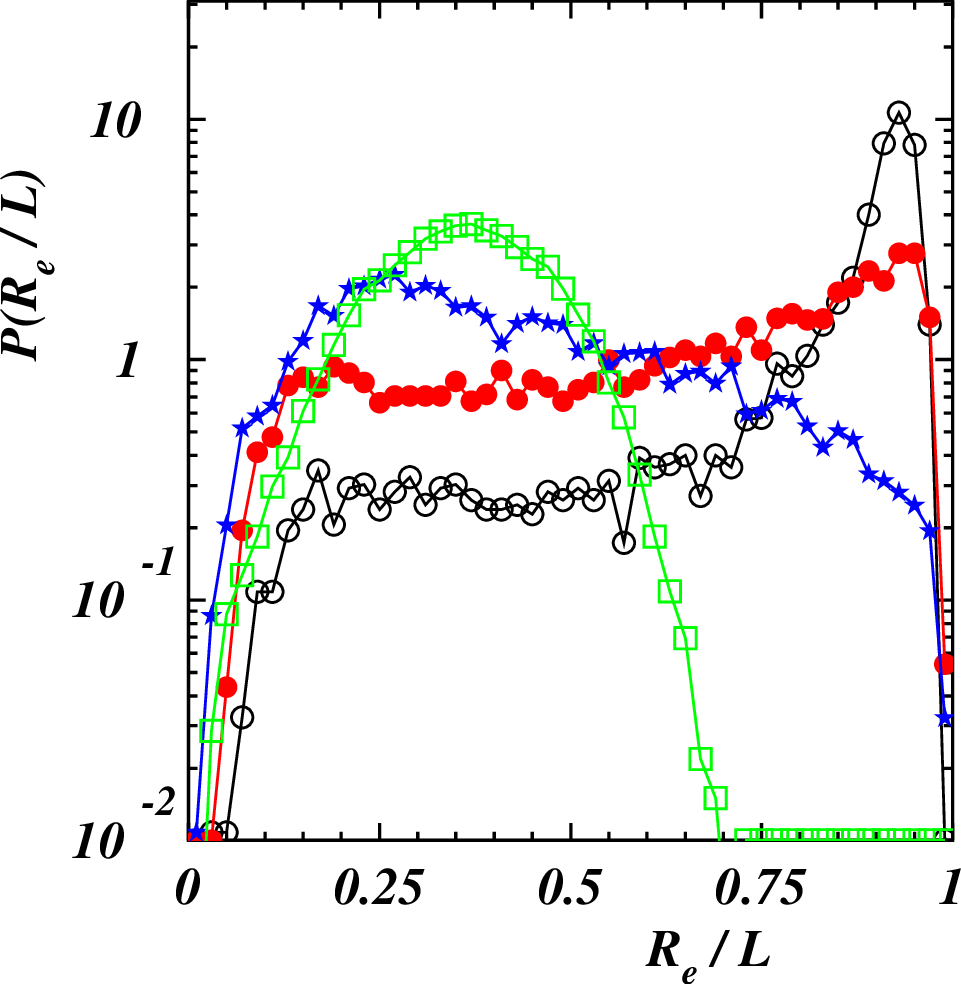}
\caption{Probability distribution functions 
of the polymer end-to-end distance
for $L_p/L=0.02$ with $Wi=0$ ($\Box$) and with $Wi=500$ for
$De=10 (\circ), 50 (\bullet), 
200 (\star)$.
\label{fig:pdflen}
}
\end{center}
\end{figure}
The behavior of the end-to-end distance in the flexible case can be analyzed 
by considering
its probability distribution function,
shown in  Fig.~\ref{fig:pdflen}.
Under equilibrium conditions, the distribution
function is characterized by a single
peak at $R_e/L \simeq 0.35$.
At $Wi/De=2.5$ the peak is less pronounced and shifted to smaller
values of $R_e$ indicating that more crumpled conformations are favored
in the intermediate strain regime. At the same time, the distribution broadens.
When increasing the strain at fixed $Wi$, small values of $R_e$ are
equally sampled and a peak appears corresponding to stretched
configurations. Further decreasing $De$ reduces the probability at smaller
values of the end-to-end distance and enhances the peak at $R_e/L \simeq 0.95$
corresponding to elongated conformations.

\begin{figure}[H]
\begin{center}
\includegraphics*[width=\textwidth,angle=0]{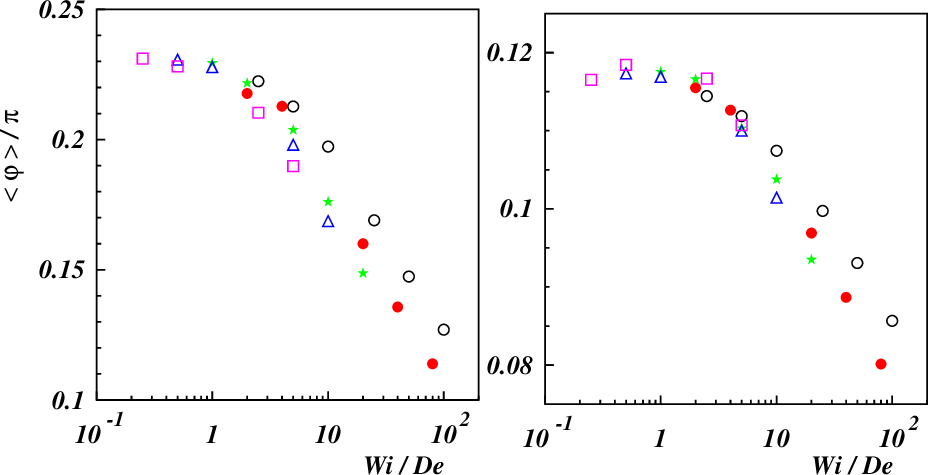}
\caption{Mean bond angles
as a function of the strain $Wi/De$
for $L_p/L=0.02$ (left), $2$ (right) for
$De=10 (\circ), 25 (\bullet), 50 (\star), 100 (\triangle), 
200 (\square)$. 
\label{fig:bond}
}
\end{center}
\end{figure}
Finally, the average bond angle 
$\lla \varphi \rra=\sum_{i=1}^{N-2}\lla \varphi_i\rra/(N-2)$ is plotted
as a function of the strain for flexible and semiflexible polymers
in Fig.~\ref{fig:bond}.
It appears that  $\lla \varphi \rra$ is very small for stiffer chains
indicating that polymers are, on the average, aligned with the flow.
The angle  $\lla \varphi \rra$ is approximately constant at small strains
and slowly diminishes when increasing the strain.
Differently, flexible filaments are characterized at equilibrium
by larger values of the
average bond angle which decreases with a faster rate 
as the strain increases. The average bond angle of flexible polymers is larger
with respect to stiffer filaments over the whole strain interval.
In both cases, a weak logarithmic dependence on strain can be observed.

\subsection{Dynamics}

\begin{figure}[H]
\begin{center}
\includegraphics*[width=0.6\textwidth,angle=0]{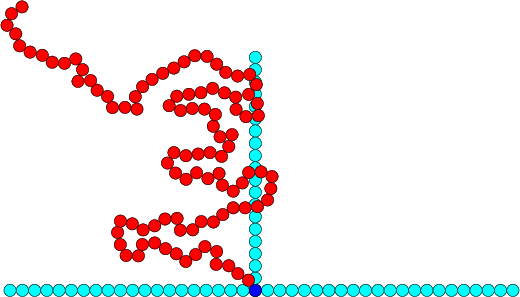}
\caption{Polymer conformation at flow inversion
  for $L_p/L=0.02$ with $Wi=500$ and $De=200$. The semi-axes of the 
reference frame are of length $20 r_0$.
\label{fig:conf}
}
\end{center}
\end{figure}

The dynamical behavior of polymer depends on the applied strain.
At high values, the filament 
moves back and forth following the external imposed flow.
A flexible chain, due its negligible bending rigidity, is enforced to
recoil at flow inversion, before to flip and elongate again. 
At smaller values of the strain, it may well happen
that the chain recoils without flipping when the flow is reverted.
An example of chain configuration at flow inversion, when $Wi/De=2.5$, 
is reported in Fig.~\ref{fig:conf}. 

\begin{figure}[H]
\begin{center}
\includegraphics*[width=\textwidth,angle=0]{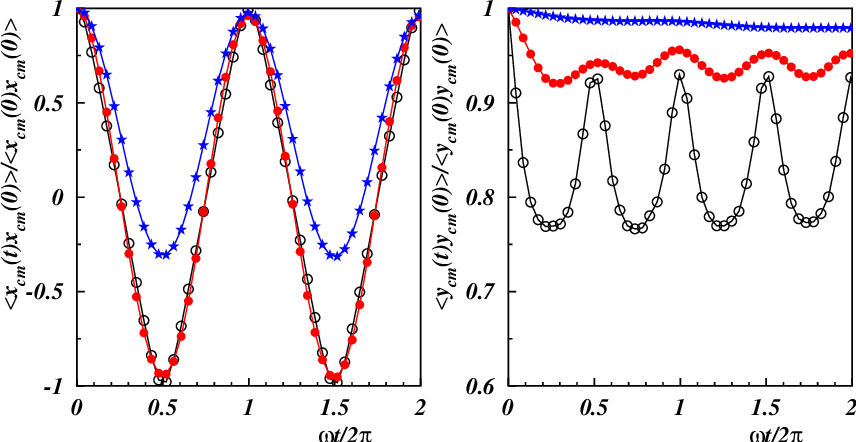}
\caption{Time auto-correlations of the center-of-mass $x$ (left) and 
$y$ (right) Cartesian coordinates
for $L_p/L=0.02$ with $Wi=500$ and 
$De=10 (\circ), 50 (\bullet), 
200 (\star)$. 
\label{fig:autocorrcm}
}
\end{center}
\end{figure}
The autocorrelation functions $\lla x_{cm}(t)x_{cm}(0)\rra$ and
$\lla y_{cm}(t)y_{cm}(0)\rra$
for the Cartesian COM coordinates $(x_{cm},y_{cm})^T$ 
are shown in Fig.~\ref{fig:autocorrcm}.
A periodic motion is clearly visible in both spatial directions
with amplitudes that are amplified by strain.
The $x$-component oscillates with the same frequency of the external flow while
the vertical component indicates a 
frequency doubling for $Wi/De \gtrsim 10$.
This behavior can be attributed to the  presence
of the reflecting wall at $y=0$ 
which pushes the polymer upwards  when  the chain flips following
the flow inversion. The occurrence of
this mechanism twice in every cycle is responsible
for the observed frequency doubling in the dynamical behavior along
the shear direction.
The negligible bending rigidity of flexible polymers
causes smaller amplitudes in the $y$-direction, with respect to stiffer chains
\cite{lamu:21}, since filaments are, on the average,
more coiled and less stretched as previously discussed. 

\begin{figure}[H] 
\begin{center}
\includegraphics*[width=0.9\columnwidth,angle=0]{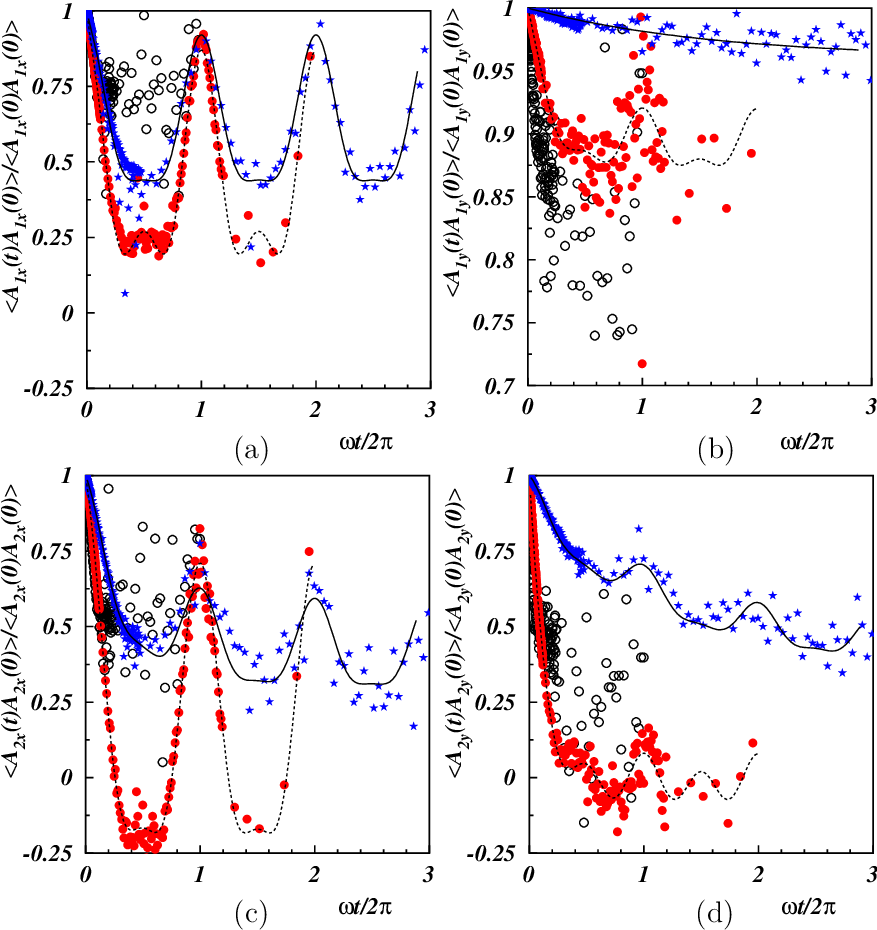}
\caption{Autocorrelation function of the mode amplitudes 
for $n=1$ (top) and $n=2$ (bottom) 
as a function of  time along the $x$- (left) and $y$-direction (right)
for $L_p/L=0.02$ with $Wi=500$ and 
$De=10 (\circ), 50 (\bullet), 
200 (\star)$. The black dashed and
solid lines are fits to the data with $De=50$ and $De=200$, respectively 
(see the text for details).
\label{fig:autocorr}
}
\end{center}
\end{figure}

In order to further characterize the internal dynamics, it may be helpful
to consider the normal mode expansion
\begin{equation}\label{modes}
{\bm r}_i = \sum_{n=1}^{N-1}{\bm A}_n(t) \sin[q_n(i-1)] \;\;\;\; i=1,\ldots,N
\end{equation}
where ${\bm A}_n=(A_{nx},A_{ny})^T$ are the normal mode amplitudes and 
$q_n=(n-1/2) \pi/(N-1)$ ($n=1,\ldots,N-1$) are the wave vectors.
It is straightforward to show that
\begin{eqnarray}\label{comp}
A_{nx}&=&\frac{2}{N-1}\sum_{i=1}^{N} r_{x,i} \sin[q_n(i-1)] ,\\
A_{ny}&=&\frac{2}{N-1}\sum_{i=1}^{N} r_{y,i} \sin[q_n(i-1)] . 
\end{eqnarray}
The  dynamics  is investigated 
by considering the mode-autocorrelation functions
$\lla A_{nx}(t)A_{nx}(0) \rra$ and $\lla A_{ny}(t)A_{ny}(0) \rra$
for the first two modes. 
The time behaviors for different strains 
are depicted in Fig.~\ref{fig:autocorr}.
Numerical data of the mode-autocorrelation functions
are fitted by using the function
\begin{equation}
  f(t)=A [\exp(-\gamma \omega t)-1] +B [\cos(\omega t)-1]
  + C [\cos(2 \omega t)-1] +1
\label{eq:autocorr}
\end{equation}
where $\gamma$ characterizes the relaxation time and
the term proportional to $C$
is considered to take into account the observed frequency
doubling.
Data with $De=50$ and $De=200$ are well fitted by Eq.~(\ref{eq:autocorr})
while this is not doable when $De=10$.
The behavior of the factor $\gamma$ depends on the mode number $n$, the
spatial direction, the strain $Wi/De$, as well as the bending rigidity. 
Along the $x$ direction, $\gamma$ decreases with increasing $n$ thus suggesting
that relaxation is faster on larger length scales along the flow direction.
The values of 
$\gamma$ along the $y$-direction does not change significantly with
$n$ and are smaller compared to the values for the $x$-direction revealing
a slower relaxation along the shear direction. Moreover, the factor
$\gamma$ increases with the ratio $Wi/De$ since the external flow
facilitates the relaxation process.
Finally, $\gamma$ is larger for flexible polymers with respect to semiflexible
ones since the applied field can easily deform chains speeding up their
relaxation. We also remark that
there is, though small, a 
contribution to the oscillations of the autocorrelation functions
from the term proportional to $C$ which indicates that the frequency
doubling observed for the motion of the COM impacts also on
the internal polymer dynamics.

\section{Conclusions}
\label{concl}

The static and dynamic behavior of filaments with vanishing small bending
rigidity have been
investigated in two dimensions under the action of an externally
imposed oscillatory linear flow. The polymer is tethered at one end on a 
wall which restricts its motion in a half-plane. 
The dynamics is numerically studied by adopting the Brownian version
of the multiparticle collision method and compared to the one of semiflexible
polymers.

The strain $Wi/De$
allows us to discriminate three regimes. When  $Wi/De \lesssim 1$, 
polymers behave as at equilibrium so that no relevant effect induced by 
the external field can be detected. At high values of strain, 
polymers follow the imposed flow alternatively
flipping from one side to the other. In this case the average deficit-length
ratio shows a power-law behavior, $\simeq (Wi/De)^{-2/3}$, as predicted for 
tethered flexible chains under steady shear. 
For intermediate values of strain such that $1 \lesssim Wi/De \lesssim 10$,
chain flipping may be unfavored due to flexibility. In this case
more coiled configurations are observed. These conformations
characterized by larger bond angles with
respect to semiflexible polymers.

The internal dynamics is examined by
investigating the behavior of the autocorrelation function of the 
center of mass. The COM motion along the flow direction is
periodic displaying the same frequency of the external shear.
Constraining the polymer in a half-plane forces the COM to move up and down 
twice per cycle. This determines
a frequency doubling in the motion along the shear direction. 
Moreover, a very small bending rigidity  
reduces the amplitude of this latter motion
and allows a faster
polymer relaxation, as observed 
by looking at the mode-autocorrelation function.

In the present study hydrodynamic effects have not been considered willing
primarily to elucidate
the role of the oscillatory flow on the polymer phenomenology. It
is well-known that hydrodynamics is able to deeply impact on the overall
dynamics, especially in the coiled state further accelerating the polymer
motion. A complete description would therefore require a numerical model
with fully resolved hydrodynamics. However, the present study has allowed
us to elucidate the relevant differences with respect to the case of
stiff filaments and will hopefully stimulate further studies of flexible
polymers in oscillatory flow fields.  

\section{Acknowledgments}

Funding from MIUR Project No. PRIN 2020/PFCXPE is acknowledged.
This work was performed under the auspices of GNFM-INdAM.





\bibliographystyle{elsarticle-num-names} 

\end{document}